\documentstyle[12pt,draft]{article}
\begin{document}
\begin{flushright}
  hep-ph/9501345
\end{flushright}

\begin{center}
\vspace{1cm}
      { \LARGE\bf Intermittency and Bose-Einstein correlations}\\
\vspace{1cm}

{\Large I.V. Andreev$^1$, M. Biyajima$^2$,
I.M. Dremin$^1$, N. Suzuki$^3$ }\\
\vspace{0.5cm}

{\normalsize  $^1$Lebedev Physical Institute, Moscow, Russia \\
 $^2$Department of Physics, Faculty of Liberal Arts,
     Shinshu University, Matsumoto 390, Japan \\
 $^3$Matsusho Gakuen Junior College, Matsumoto 390-12, Japan\\ }
\end{center}
\begin{abstract}
The role of Bose-Einstein correlations in a widely discussed
intermittency
phenomenon is reviewed. In particular, it is shown that particle
correlations of different origin are better displayed when analysed
as functions of
appropriately chosen variables. Correspondingly, if the shape
of Bose-Einstein contribution is chosen to be Gaussian in
3-momentum transferred, it provides
the power-like law in 4-momentum squared and is smeared out in
(pseudo)rapidity.
\end{abstract}

             \section { Introductory survey and definitions.}

     The  study  of  fluctuations  and  correlations  in  hadron
production at high energies has found considerable  interest  in
recent years. The $q$-particle inclusive densities $\rho_{q}(p_{1},
\ldots,p_{q})$  or  rather
the factorial moments $\langle n(n-1)\ldots (n-q+1)\rangle $
estimated in  different  phase  space
regions $\delta $ were studied in a variety of reactions
ranging  from
$e^{+}e^{-}$ to nucleus-nucleus scattering.

     The concept of intermittency has been introduced  in  order
to describe enhanced fluctuations observed for individual events
in the density distributions of hadrons (for a recent review see
\cite{1}). Originally the  definition  of  intermittency  was  the
strict power laws of the normalized factorial moments taken as a
function of bin size \cite{2} at small bin sizes (self-similarity of
the moments). Later  on  it  has  become  customary  by  calling
intermittency  any  increase  of  the  factorial  moments   with
decreasing phase  space  intervals  without  regard  to  scaling
behaviour.  The  extended   version   indicates   just   positive
correlations in momentum space which increase at  smaller  bins.
Such an increase of  factorial  moments  demonstrates  that  the
multiplicity distributions become wider in small phase bins i.e.
fluctuations are stronger.

     Being important by itself, it shadows a crucial feature  of
the original proposal, namely, the search for self-similarity in
the processes which could be  related  \cite{3}  to  the  fractal
structure  of  the  pattern  of  particle   locations   in   the
three-dimensional  phase  space.  Therefore   using   the   term
"intermittency" in a wide sense here, we still keep in mind that
the scaling regime is of primary interest. It gave name  to  the
whole effect when was first  proposed  in  study  of  turbulence
\cite{4}.

     We shall discuss some possible sources of correlations. Our
main concern is to show that they  contribute  differently  when
the  various  projections  of  multiparticle  phase  space   are
considered. The proper choice of the variable can emphasize  the
particular correlation and, vice versa, conceal the  contribution
of others. Therefore, by choosing { \it different} variables  we
 study the correlations of {\it different}  origin  weighted
with {\it different} weights.

Various explanations with different physical origin  of
correlations have been proposed. We shall discuss them in more
detail in Section 3, but here let us just mention some of  them.
The analogy to turbulence has led in the  original  proposal  to
the   term   "intermittency"   and   to   the   phenomenological
multiplicative cascade model of the phenomenon.
On  more  strict
grounds, it could be related to the {\it parton}  shower  in
quantum chromodynamics (QCD) as described in Section 3.5.
The observed phenomenon of {\it hadron} jets  by
itself indicates strong positive correlations. Thus, the scaling
regime could appear also at the stage of transition from partons
(quarks, gluons) to the observable particles  (mostly  hadrons).
Therefore the ideas of phase  transition  have  been  elaborated
too. Even a more trivial dynamical  reason  could  be  connected
with abundant production of resonances which, surely,  imply  the
correlation  of  relative  energies  of  final  particles.  Some
unknown sources of completely new dynamics have been looked  for
(e.g. stochastic dynamics, instabilities etc.).

     Beside  those  "dynamical"   effects   there   exists   the
well-known symmetry property of the interaction which  necessarily
contributes  to  the  enhancement  of  correlations  in   small
phase-space volumes. We mean Bose-Einstein correlations  due  to
the symmetrization of the wave functions of identical particles.

In this review, we shall deal mostly with the  problem  of  what
and how the Bose-Einstein effect  contributes  to  the  observed
increase  of  moments  of  multiplicity  distributions  in  ever
smaller phase space regions when various variables are chosen to
tag the phase space bin.  Other  topics,  mentioned  above,
will be considered to the extent  they  are  necessary  for
clarification of some issues related to our main purpose.

     Bose-Einstein (BE) statistics leads  to  specific  positive
correlations  of  identical  bosons  (in   multiple   production
processes those are mainly like-charged pions). BE  correlations
are  stronger  for  particles  having  smaller   difference   of
three-dimensional  momenta.   So   BE   correlations   lead   to
intermittency effect (in the broad sense of the word).
Indeed  it was first considered in \cite{5a}, and
was asserted a few years ago \cite{5b,6,7} that BE correlations
may  be responsible for the intermittency  effect.  However
ineffective
methods of the data analysis did not give a possibility to reach
convincing conclusions that time.

     Early investigation of the intermittency  was  concentrated
on the study  of  one-dimensional  rapidity  dependence  of  the
traditional factorial moments in decreasing rapidity  intervals.
Then  it  was  realized  that intermittency effect in
multiparticle
production, if any, would naturally occur also in two and  three
dimensions when two  or  all  three  components  of  momenta  of
outgoing particles are registered.

     Furthermore  high-order  inclusive  density  $\rho_{q}
(p_1 ,...,p_q )$  contains contributions from lower order
correlations. Being interested in
genuine multiparticle correlations one  is  led  to  investigate
(connected)  correlation   functions  $C_{q}(p_1 ,...,p_q )$
with   lower
order correlations subtracted (the correlation  function  $C_q$
vanishes when a subgroup of $m<q$  particles  is
statistically  independent
from the other $q-m$ particles).

     The normalized factorial  moment  estimated  in  the  phase
space region $\delta $ is defined as
\begin{eqnarray}
F_q(\delta )=\frac {F_{q}^{(u)}(\delta )}{\langle n\rangle ^q}
=\frac {1}{\langle n\rangle ^q}\int_{\delta }dp_1 \ldots
\int_{\delta }dp_q \rho _{q}(p_1 ,\ldots,p_q)={} \nonumber \\
{}=\frac {\langle n(n-1)\ldots (n-q+1)\rangle }
{\langle n\rangle ^q},
\label{1.1}
\end{eqnarray}
where $F_{q}^{(u)}$ are called unnormalized moments. The
normalization is  chosen  so  that  at integer ranks $q$
one gets $F_q \equiv 1$ for Poisson
distribution. In Eq.~(\ref{1.1}) $dp_j$ is the differential
momentum space volume
\begin{equation}
dp_j =d^{3}p_{j}/2E_{j}(2\pi )^3       \label{1.2}
\end{equation}
or any other differential interval of interest. Here $\delta $
will  be used as a general notation for a  selected  phase
space  volume which   differs   for   different   selection
procedures.   In one-dimensional  analysis  this  is  usually
the  interval   of rapidity $\delta y$ or pseudorapidity $\delta
\eta $ where rapidity is defined as
$y=\frac {1}{2}\ln \frac {E+p_l}{E-p_l}$
and pseudorapidity as $\eta =\frac {1}{2}\ln \frac {p+p_l}{p-p_l}=
\ln \tan \frac {\theta }{2}$ with $E, p, p_l ,\theta $ being the
 energy, the momentum,
the longitudinal momentum and the angle of particle emission.

          In practice, an averaging over cells of the original
phase space is performed, so  that  traditional  normalized
factorial moments in $d$ dimensions are determined as
\begin{equation}
F_{q}(\delta )=\frac {1}{M^d}\sum_{m=1}^{M^d}F_{q,m}(\delta ),
\;\;\;d=1, 2, 3,    \label{1.3}
\end{equation}
where the sum runs over $M^d \; dq$-dimensional  boxes  having
the  same
size. Self-similarity of the moments $F_{q}(\delta )$ taken as a
function  of momentum cell size suggests a power law behaviour
\begin{equation}
F_{q}(\delta )=\left (\frac {\Delta }{\delta }\right
)^{\phi _q}F_{q}(\Delta ),
\label{1.4}
\end{equation}
where $\phi _q >0$ are known as intermittency indices.

     General interrelation of the inclusive densities $\rho _q$
and the correlation functions  $C_q$  is  provided  through
the  inclusive generating functional
\begin{equation}
G[Z]=\sum_{q=0}^{\infty }\frac {1}{q!}\prod_{j=1}^{q}
\int dp_{j}\rho _{q}
(p_1 ,\ldots,p_q )Z(p_1)\ldots Z(p_q) ,  \label{1.5}
\end{equation}
where $Z(p_j)$ are auxiliary functions. The  inclusive  densities
 are then given by differentiation
\begin{equation}
\rho_{q}(p_1,\ldots,p_q)=\frac {\delta ^{q}G}
{\delta Z(p_1)\ldots \delta Z(p_q)}
\vert _{Z=0} .    \label{1.6}
\end{equation}
The correlation functions $C_q$ are defined through the  generating
functional $G$ in the following way (cluster expansion):
\begin{equation}
G[Z]=\exp (\int dp_{1}\rho _{1}(p)Z(p)+\sum _{q=2}^{\infty }
\frac {1}{q!}
\prod _{j=1}^{q}\int dp_{j}C_{q}(p_1,\ldots ,p_q)Z(p_1)
\ldots Z(p_q))
\label{1.7}
\end{equation}
being "the exponents" of the inclusive densities,
\begin{equation}
C_{q}(p_1,\ldots ,p_q)=\frac {\delta ^{q}\ln G}
{\delta Z(p_1)\ldots \delta
Z(p_q)}
\vert _{Z=0} .  \label{1.8}
\end{equation}

     A comparison of (\ref{1.5}) and (\ref{1.7}) leads to
relationships
\begin{eqnarray}
\rho _{2}(p_1, p_2)=\rho _{1}(p_1)\rho _{1}(p_2)
+C_{2}(p_1, p_2) ,\\
\rho _{3}(p_1, p_2, p_3)=\rho _{1}(p_1)\rho _{1}(p_2)
\rho _{1}(p_3)+ \nonumber
\\ \sum _{i\neq j\neq k; i=1}^{3}\rho _{1}(p_i)C_{2}(p_j, p_k)+
C_{3}(p_1, p_2,
p_3)  \label{1.9} \end{eqnarray} etc, or the inverse ones:
\begin{eqnarray}
C_{2}(p_1, p_2)=\rho _{2}(p_1, p_2)-\rho _{1}(p_1)
\rho _{1}(p_2) ,\\
C_{3}(p_1, p_2, p_3)=\rho _{3}(p_1, p_2, p_3)-
\sum _{i\neq j\neq k; i=1}^{3}\rho_{1}
(p_i)\rho _{2}(p_j, p_k) \nonumber  \\ +2\rho _{1}(p_1)
\rho _{1}(p_2)\rho
_{1}(p_3) \label{1.10} \end{eqnarray} etc. Let us note that the
correlation functions have  the  above mentioned physical
interpretation  if  the  number
of  produced particles exceeds (not clear a priori to what
extent) the  order of the correlation function.

          Integrated (and normalized) correlation functions
are known as cumulant moments
\begin{equation}
K_{q}(\delta )=\frac {K_{q}^{(u)}(\delta )}{\langle n\rangle ^q}=
\frac {1}{\langle n\rangle ^q}\int dp_1 \ldots
\int dp_{q}C_{q}(p_1,\ldots,p_q)
{}.
\label{1.11}
\end{equation}
Cumulants of Poisson distribution are identically equal to zero.

The generating function of the moments arises if  one puts the
functions $Z(p_j)$ in generating functional (\ref{1.5}) to be
a constant $z$,
\begin{equation}
g(z)=\sum _{q=0}^{\infty }\frac {1}{q!}F_{q}^{(u)}z^q
=\exp (\langle n\rangle
z+\sum _{q=2}^{\infty }\frac {1}{q!}K_{q}^{(u)}z^q)\\
=\sum _{n=o}^{\infty }
P_{n}(1+z)^n .  \label{1.12}
\end{equation}
The moments and particle number distribution $P_n$ can be found
by differentiation of the generating function:
\begin{eqnarray}
\frac {d^{q}g(z)}{dz^q}\vert _{z=0}=F_{q}^{(u)}=\langle n
\rangle ^{q}F_q=
\langle n(n-1)\ldots (n-q+1)\rangle , \label{1.13}\\
\frac {d^{q}\ln g(z)}{dz^q}\vert _{z=0}=K_{q}^{(u)}=
\langle n\rangle ^{q}K_q ,
\label{1.14}  \\
\frac {d^{n}g(z)}{dz^n}\vert _{z=-1}=P_{n}n! . \label{1.15}
\end{eqnarray}
     Relationships  between  factorial  moments   and   cumulant
moments follow from (\ref{1.12}) or from (\ref{1.9})
and (\ref{1.10}):
\begin{eqnarray}
F_{2}=1+K_2 , \\
F_{3}=1+3K_2 +K_3 ,   \\
F_{4}=1+6K_{2}+3K_{2}^{2}+4K_{3}+K_4 , \ldots         \label{1.16}
\end{eqnarray}
for a fixed cell in momentum space.  An  averaging  over  cells,
such as that in (\ref{1.3}),
\begin{equation}
K_{q}=\frac {1}{M^d}\sum _{m=1}^{M^d}K_{q,m}(\delta ) \label{1.17}
\end{equation}
requires the corresponding averaging of Eqs. (\ref{1.16}).

     Let us return to  the  intermittency  study.  The  cumulant
moments    (\ref{1.17}), representing genuine  multiparticle
correlations, suffer from arbitrary binning and low  statistics.
The accuracy in the measurement of the factorial moments is also
unsatisfactory  for  small  bin   sizes.   That   has   led   to
investigation of more general density and correlation  integrals
which give the possibility to use the available statistics in an
optimal way and to perform the correlation analysis in any
convenient variable.

     The general density integral is defined as
\begin{equation}
F_{q}^{\Omega }(\delta )=\frac {1}{N_q}\prod _{j=1}^{q}
\int dp_{j}\rho _{q}
(p_1,\ldots ,p_q)\Omega _{q}(\delta ;p_1,\ldots ,p_q) ,
\label{1.18}
\end{equation}
and the correlation integral as
\begin{equation}
K_{q}^{\Omega }=\frac {1}{N_q}\prod _{j=1}^{q}
\int dp_{j}C_{q}(p_1,\ldots ,p_q)
\Omega _{q}(\delta ;p_1,\ldots ,p_q) , \label{1.19}
\end{equation}
where the "window function" $\Omega _{q}$ is determined to
be nonzero
in some prescribed interval of the $q$-particle phase space.
The normalization  to  uncorrelated  background  is
suggested   in
(\ref{1.18}), (\ref{1.19}):
\begin{equation}
N_q=\prod _{j=1}^{q}\int dp_{j}\rho _{j}(p_j)
\Omega _{q}(\delta ;p_1,\ldots,p_j)
{}.
 \label{1.20}
\end{equation}
     The  general  definitions   (\ref{1.18}),   (\ref{1.19})
give   a possibility not to split  the  $q$-particle  phase  space
in  an artificial way (as it was the case in  Eqs.  (\ref{1.3}),
(\ref{1.17}))
ensuring better statistics. This in turn gives a possibility  to
measure $F_{q}^{\Omega }$ and $K_{q}^{\Omega }$ for like-charged
and unlike-charged  particles
separately providing direct evidence for the BE correlations.

               \section  { Experimental survey.}

     The extensive review  of  experimental  data  is  given  in
\cite{1}. Here, we mention some typical findings which  help  us
exemplify  our  treatment,  especially,   those   which   reveal
BE-contributions.

     \subsection { One-dimensional moments (rapidity variable).}

      Consider first the  one-dimensional  analysis  in  rapidity
variable.  Factorial   moments   (\ref{1.3})   for   like-charged
particles were measured in $\pi ^{+}p$ and $K^{+}p$ interactions
 at  250  GeV/c by  NA22  Collaboration  \cite{8}.  They  are
characterized  by intermittency indices $\phi _q$ according to
the fit \begin{equation}
F_{q}(\delta y)=a_{q}(\delta y)^{-\phi _{q}} .   \label{2.1}
\end{equation}
Though the  bin  size  dependence  turns  out  to  be  far  from
power-like in the whole  interval  measured,  but  it  could  be
fitted by power laws separately at large ($\delta >1$) and small
($\delta <1$)  bins.
We remind that the latter region was suspected for such a law in
original proposal (see Table 1 and Fig. 1).

\begin{table}
\caption { Intermittency indices for  all  charged  particles  and
         negatives only (for small bins )}
\begin{tabular}{l|l|l|l|l}  \hline
           &   $\phi _2$     &    $\phi _3$    &   $\phi _4$   &
$\phi _5$
\\ \hline
all charged& $0.008\pm 0.002$&$ 0.043\pm 0.006$&$ 0.16\pm
0.02$&$ 0.39\pm 0.06$
\\ \hline
negatives  & $0.007\pm 0.003$&  $0.06\pm 0.02$ & $0.29\pm 0.06$&
\\
only       &                 &                 &               &
\\ \hline
\end{tabular}
\end{table}
The values of the moments are  smaller  but  the  increase  with
decreasing bin sizes is stronger when a  like-charge  sample  is
used instead of all charged  sample.  To  what  extent  the  BE
correlations (presumably dominating the correlation of like-sign
particles) are responsible for the  full  intermittency  is  not
clear from the data in the rapidity variable.  Anyhow  they  are
not dominating.

     Analogous analysis of the factorial moments  was  performed
in $p\bar p$ collisions at $s^{1/2}=630$ GeV by UA1 Collaboration
\cite{9}, see  Fig. 2. The authors conclude that the BE  effect
is  weak  in  this variable.
(Their early data of intermittency were analysed by means of the
negative binomial distibution and the pure birth stochastic
theory in Ref.\cite{9b}.)\\

 \subsection { Higher-dimensional analysis of the factorial
moments.}

     The intermittency effect is  more  pronounced  in  two  and
especially in three dimensions. The impressive  results  on  the
second factorial moment were reported in $\mu N$ interactions at
280 GeV/c  by  NA9  Collaboration  \cite{10}.  The  moments
$F_2$  for unlike-charged and for negative particles were measured
in  one and  three  dimensions.  The  number  of  boxes  is  $M$
and  $M^3$ correspondingly. The results are given in Fig.  3.
The  authors claim that the strong intermittency signal
observed  in  $F_{2}^{--}$  in three  dimensions  has  to  be
attributed  exclusively  to   BE
correlations since such a signal is  not  present in  $F_{2}^{+-}$.
This conclusion is supported by the fact that the  Lund  model (not
containing BE correlations) is in rough agreement with the  data
for $F_{2}^{+-}$ but in complete disagreement for $F_{2}^{--}$.
In one  (rapidity)
dimension  the  effects  are   much   weaker.   Nevertheless   a
considerable difference between data and the Lund prediction  is
seen for $F_{2}^{--}$ in one dimension too whereas for
$F_{2}^{+-}$ the difference is
small.  This  also  supports   the   noticeable role of   BE
correlations.

     Similar results  for  $F_{2}(\delta )$  in  three  dimensions
were  also presented in  $\pi ^{+}p$  and  $K^{+}p$  interactions
by  NA22  Collaboration
\cite{8}. The box volume dependence was fitted with the form
\begin{equation}
K_{2}(\delta )=F_{2}(\delta )-1=c+a\delta ^{-b} , \;\; \delta
=1/M^3 .\label{2.2}
\end{equation}
A striking difference for unlike-  and  like-charged  pairs  was
found. While the (~+~--~) pairs are dominated by long-range
correlations
(large $c$), these are smaller or absent  in  the  case
(--~--)  and
(+~+). Correspondingly, the parameter $a$ is compatible with  zero
for (+~--), but relatively large for (--~--)  and  (+~+).
This  again supports  BE  interpretation  of   the   intermittency
in   three dimensions.

     \subsection { Density integrals. Three dimensional analysis. }

     For higher orders ($q\geq 3$) the traditional  normalized
factorial moments have large statistical errors when evaluated
for  small bin sizes. So  nowadays  the  general  correlation
analysis  is usually performed (see Eqs. (\ref{1.18})-(\ref{1.20}))
permitting not to split the phase space interval under
consideration. Using
the density and correlation integral method, one must  introduce
the distance $\delta _{ij}$ for each  pair  of  particles  and
confine  the
permissible $q$-particle phase space. Usually the  "distance"  is
defined as four-momentum difference squared,
$\delta _{ij}=Q_{ij}^{2}=(p_{i}-p_{j})^2$. The permissible
phase space is most commonly confined either  by  GHP  condition
\cite{11}, $\delta _{ij}\leq \delta $ for all $i,j\leq q$ or by
so called star topology,  $\delta _{1j}\leq \delta $  for
$j=2,\ldots,q$, or by snake topology, $\delta _{j,j+1}<\delta
$ for $j=1,\ldots,q-1$.

     A  comparison  of  GHP-integral   with   the   conventional
normalized  factorial  moments  was  performed  in  $\pi ^{+}p$
and   $K^{+}p$ interactions  at  250   GeV/c   by   NA22
Collaboration.   The three-dimensional "distance" between two
particles $i$ and $j$ was defined using $y, \varphi $ and $p_{t}$
variables in a rather specific way as
\begin{equation}
d_{ij}=\max (\vert y_i-y_j \vert , \vert \varphi _i -
\varphi _j \vert ,
 \vert p_{t,i}-p_{t,j} \vert )^3 .   \label{2.3}
\end{equation}
The size $\delta $ of $q$-tuple was defined by the smallest  box
volume that encloses the $q$-tuple. The  determination  of  the
density integrals can now be compared with moving around a  box
in  the
full phase space under consideration and counting the number  of
the $q$-tuples fitted into the box.

     In Fig. 4 conventional moments $F_2$  and  $F_3$  are
compared  to  the density integral version $F_{2}^{GHP}$ and
$F_{3}^{GHP}$. As anticipated, in  Fig.  4
one indeed observes  that  statistical  errors  in  the
$F_{q}^{GHP}$  are
strongly reduced. This, in principle, allows the analysis to be
carried down in much smaller box volumes. It, furthermore,
allows  a
comparison of different charge combinations.

     For the second-order integral $F_{2}^{GHP}(\delta )$ the
fit (\ref{2.2})
was used as it
was the case for conventional factorial moments considered in
the previous
 section. The results are shown  in
Table 2 (we omitted the constant $b$ when the constant  $a$  was
found to be compatible with zero) being qualitatively the same as
for ordinary three-dimensional factorial moments.

\begin{table}
\begin{center}
\caption{Results of fits (26) to the data on $F_{2}^{GHP}$}
\medskip
\begin{tabular}{l|l|l|l}   \hline
     &    unlike charged   & negatives only      &   positives
only \\ \hline
$ a$   &  $ 0.0006\pm 0.0009$& $0.0115\pm 0.0003$  &    $ 0.04\pm
0.02$ \\
\hline
$ b  $ &                     &  $0.469\pm 0.004$   &    $ 0.37\pm
0.06$  \\
\hline
 $c $  &  $ 0.380\pm 0.006$  &       $ 0 $         &    $ 0.08\pm
0.03$   \\
\hline
 \end{tabular}
\end{center}
\end{table}
The  above  results  again  support  the
conclusion  that   the intermittency in three dimensions is
strongly enhanced
due to BE correlations.

     For the higher order GHP  density  integrals  the  modified
power law assumption
\begin{equation}
\ln F_{q}^{GHP}(\delta )=a_{q}+\frac {\alpha _{q}}{\alpha _{2}}
F_{2}(\delta )
\label{2.4}
\end{equation}
can be fitted to the data though (\ref{2.4}) only holds
approximately.
The results are presented in Table 3.

\begin{table}
\begin{center}
  \caption{ The slopes $\alpha _{q}/\alpha _{2}$ obtained by
  fitting (28) to the data}
\begin{tabular}{l|l|l|l}   \hline
                      &  all charged   & negatives only    &
positives only \\
\hline
$\alpha _3/\alpha _2$ &$3.81\pm 0.09$  &    $4.3\pm 0.2$   &
    $5.3\pm 0.2$ \\
$\alpha _4/\alpha _2$ &  $8.0\pm 0.03$ &    $9.4\pm 0.5$   &
   $15.0\pm 0.5$ \\
$\alpha _5/\alpha _2$ & $11.8\pm 0.3$  &   $13\pm 0.1$     &
   $24\pm 1   $ \\
\hline
\end{tabular}
\end{center}
\end{table}
As can  be  seen  from  the  Table,  the
growth  in  decreasing three-dimensional phase-space volumes is
faster for
higher order density integrals and for like-charged particles.

Let us mention
here the analysis \cite{12} done by the same NA22  Collaboration
using the opening angle $\theta _{ij}$ between two particles
\begin{equation}
\theta _{ij}=\arccos ({\bf p}_{i}{\bf p}_{j}/\vert {\bf p}_{i}
\vert \vert
{\bf p}_{j}\vert ) \label{2.4.1}
\end{equation}
with ${\bf p}_{i}$ and ${\bf p}_{j}$ being the three-momenta
of particles
$i$  and  $j$.
An angular distance measure  for  more  than  two  particles  is
defined as the maximal relative  angle  between  all  the  pairs
chosen. Therefore, the numerator of the factorial moment of rank
$q$ is determined by counting, for  each  event,  the  number  of
$q$-tuples that have a pairwise angular opening  smaller  than  a
given value and then averaging over all events.

        The fitted values of intermittency indices appeared to be
very low (sometimes negative)  and  strongly  dependent  on  the
production angle of the particles. We argue in Section 3.5  that
such  an   analysis   reminds   of  two-dimensional analysis
in   relative
pseudorapidity and azimuthal angle but with the bin  size  which
depends on the production angle. Therefore it is not very useful
for comparison with analytical  calculations  for  parton  jets.
Nevertheless, it is possible to  compare  the  experimental
findings with  the  FRITIOF  Monte-Carlo  model. It shows that BE
correlations should be incorporated into the model  to  get  the
agreement.

      \subsection { $Q^2$-analysis of the density integrals.}

      NA22  Collaboration  has  also  presented the data on
$Q^2$-dependent GHP density integrals, that is the integration
over particle densities $\rho _q$ was confined by the factor
\begin{equation}
\Omega _{q}^{GHP}(Q^2)=\prod _{i<j}^{q} \Theta (Q^{2}-Q_{ij}^{2}) .
 \label{2.5}
\end{equation}
The data are shown in Fig. 5 and Table 4 where the parameters
of the power law fit
\begin{equation}
F_{q}^{GHP}(Q^2)=a_{q}(Q^{2})^{-\phi _{q}}  \label{2.6}
\end{equation}
are given. One can see that evaluation of the GHP  integral  for
different charge  combinations  yields  effective  intermittency
indices  a  factor  1.2  ($\phi _{5}$)  to  1.6  ($\phi _{2}$)
larger for   the
negatives-only  sample  than  for   all-charged   sample.   This
indicates the important role of BE correlations as displayed  on
$Q^2$-scale though more definitive conclusions can be hardly
 reached from the data.

\begin{table}
\begin{center}
     \caption{ Results of fits to the data presented in Fig.5
               according to Eq.(31). }
\begin{tabular}{l|l|l|l}   \hline
         &   all charged      &   negatives only  &   positives
only \\ \hline
$a_2 $    &  $ 1.219\pm 0.003$   &  $ 1.131\pm 0.002 $ & $
  1.026\pm 0.002$ \\
$\phi _2$ &  $ 0.051\pm 0.001 $  & $  0.081\pm 0.001$  &  $
 0.067\pm 0.001$ \\
\hline
$a_3   $  & $  1.751\pm 0.007$   &  $ 1.38\pm 0.01$    &  $
 1.15\pm 0.05$   \\
$\phi _3$ &  $ 0.177\pm 0.002$   &   $0.253\pm 0.004$  & $
  0.227\pm 0.003$ \\
\hline
$a_{4}  $ &   $2.90\pm 0.02 $    & $  1.88\pm 0.02  $  &  $
 1.41\pm 0.01$  \\
$\phi _4$ & $  0.358\pm 0.006 $  &  $ 0.47\pm 0.01  $  &  $
 0.45\pm 0.01$ \\
\hline
$a_5     $&  $ 5.45\pm 0.08$     & $  2.78\pm 0.07 $   &  $
 1.89\pm 0.04$  \\
$\phi _5$ &  $ 0.56\pm 0.01 $    &  $ 0.66\pm 0.02 $   &  $
 0.66\pm 0.02$ \\
\hline
\end{tabular}
\end{center}
\end{table}
     Let us note in this connection that
     $p_t$-dependence  of  the second and third order GHP
density integrals
taken as a function of $Q^2$ appears to be opposite for
all-charged and
negative-only particles in this experiment, see Table 5.

\begin{table}
\begin{center}
     \caption{ Intermittency indices for different $p_t$ regions}
\begin{tabular}{l|l|l|l}   \hline
                &           & $p_t < 0.15\ $GeV/c  & $p_t>0.15
\ $GeV/c\\ \hline
all-charged     &  $\phi _2$ &   $ 0.046\pm 0.002$   &
$0.032\pm 0.001$ \\
                &  $\phi _3$ &   $ 0.136\pm 0.004$   &
$0.107\pm 0.003$ \\
\hline
negatives-only  &  $\phi _2$ &    $0.053\pm 0.002$   &
$ 0.081\pm 0.001$ \\
                &  $\phi _3$ &   $ 0.17 \pm 0.01$   &
$ 0.269\pm 0.006$ \\
\hline
\end{tabular}
\end{center}
\end{table}
According to the Table, the intermittency
effect  is  weaker at low $p_t$ than at high $p_t$  for negatives.
On  the contrary, intermittency is stronger when small $p_t$
particles  are
selected from all charged. This hampers easy  interpretation  of
the all-charged  data  only,  without  proper  analysis  of  the
different charge combinations.

The related results on density integrals  $F_{2}(Q^2)$  and
$F_{3}(Q^2)$  were
presented in $\mu N$ interaction at 490 GeV/c,  using  data
from  the  E665
experiment at the Tevatron of Fermilab \cite{13}. This time  the
star topology was used to confine the $q$-particle  phase  space,
that is the confining factor was taken in the form
\begin{equation}
\Omega _{q}(Q^2)=\prod _{j=2}^{q} \Theta (Q^2 -Q_{1j}^2 ) .
\label{2.7}
\end{equation}
The Fig. 6 shows a log-log plot of  the  density  integrals  for
different charge combinations. $F_{2}(Q^2)$ rises more steeply
with $1/Q^2$ for
(~--~--~) than for (~+~--~) pairs. The same is  true  for
$F_{3}(Q^2)$  for
(~--~--~--~) triplets as compared to ($ccc$) triplets.
This different behaviour
indicates important role of BE  correlations  between  like-sign
particles.

     No significant energy $W$-dependence of  the  $F_2$  slope
was observed for $Q^{2}\geq $ 0.01 GeV$^2$ ; for smaller $Q^2$
the $F_2$ slope  of
(~--~--~) pairs seems to be somewhat larger for  the  high-$W$
sample.  In fact, the slopes $d\ln F_2/d\ln (1/Q^2)$ of the
$F_2$ integrals show close
agreement for NA9 and E665 experiments \cite{10,13} in spite of
the  somewhat
different $\langle W\rangle $ values and the experimental
differences of the
two experiments.

    $ Q^2$-analysis of the density integrals was also performed in
$p\bar p$ reaction at $s^{1/2}=630$  GeV by UA1  Collaboration
\cite{9}.
 The  authors
however used quite another definition to confine the permissible
$q$-particle phase space,
\begin{equation}
\sum _{j_{1}<j_{2}}^{q}q^{2}_{j_{1},j_{2}}<Q^2    \label{2.8}
\end{equation}
proposed  in Ref. \cite{14},  which  is  much   stronger   than
Eqs.(\ref{2.5}),(\ref{2.7}). The results are given in Fig. 7 for
all-charged and
like-sign particles up to order $q$=5. The $Q^2$-dependence  of
 the density integrals is close to linear one in a log-log plot.
 The corresponding intermittency indices are listed in
the Table 6.

\begin{table}
\begin{center}
\caption{ The results of fitting of normalized density integrals
         as  functions of $Q^2$ to a power law in $p\bar p$
        interactions          at $s^{1/2}=630$ GeV}
\begin{tabular}{l|l|l|l|l}   \hline
   slope   &    $\phi _2$     &  $\phi _3$   & $
\phi _4$    & $\phi _5$ \\
parameters &                  &              &
              &          \\ \hline
all charged&$0.0348\pm 0.0006$&$0.078\pm 0.001
$&$0.213\pm 0.004$&$0.338\pm
0.019$\\
particles  &                  &              &
&           \\
\hline
like-sign  &$0.0522\pm 0.0009$ &$0.147\pm 0.001$&$0.443\pm 0.01
$ &$0.855\pm
0.051$\\
particles  &                  &              &
&          \\ \hline
\end{tabular}
\end{center}
\end{table}
The comparison in Fig. 7 and Table 6 shows
once again that the slope parameters $\phi _q$ are bigger for
like-sign
particles than  for all charged particles and the condition
\begin{equation}
\phi _{q}(like-sign)\approx 2\phi _{q}(all)    \label{2.9}
\end{equation}
is fulfilled approximately in the $Q^2$-representation  with
the confinement condition (\ref{2.8}). Let us note, that the
relationship (\ref{2.9}) was suggested in Ref. \cite{6}  as
an  indication  on  BE
origin of the intermittency effect (though it was advocated
for rapidity variable in \cite{6}).

     At the same time only small differences  between  different
charge combinations are visible in the  pseudorapidity  analysis
of the  same  UA1  data,  see  Fig.  2.  The  Figures  2  and  7
demonstrate  that  the  manifestation  of  BE  correlations   is
strongly dependent on the variable used.

           \subsection { Second order correlation function.}

     A detailed study of the second order normalized correlation
functions  $C_{2}(Q^2)$  (sometimes  known  as  differential
correlation integral) was undertaken in $p\bar p$ interaction by
UA1  Collaboration
\cite{9}. Fig. 8b shows a  comparison  of  the  samples  of  the
like-sign  pairs  with  unlike-sign  pairs   and   all   charged
particles.  In  $Q^2$-variable  one   observes a dominance of
unlike-sign pair correlation for  0.03 GeV$^2\leq Q^2\leq $1
GeV$^2$  which  is  at  least
partly due to resonances and particle decay  (i.e.  there
is  a broad peak at $Q^2\approx $0.5 GeV$^2$ which is due to
$\rho $-meson decays
and  a  peak  at  $Q^2\approx $0.17 GeV$^2$
which is due to $K_{S}^{0}$ decays). However at small
$Q^2\leq $0.03 GeV$^2$ this function is
nearly constant. Contrarily, the like-sign particle  correlation
function  is  rising  very  rapidly  at  very  small  $Q^2$
up  to $Q^2=0.001$ GeV$^2$ exhibiting approximately power-law
behaviour.

     A comparison  with  the  same  analysis  in  pseudorapidity
variable $\delta \eta $ (Fig. 8a) shows once more the
significance of  the  choice of the
proper  variable  in  correlation   analysis:   the   two
body correlation function of all charged particles  is
dominated  by unlike-sign  particle  correlations  when
analysed  in  $\delta \eta $  but
dominated by the like-sign correlation function when analysed at
very small $Q^2$. Similar results on $C_{2}(Q^2)$ were reported
in $\pi ^{+}p$ and  $K^{+}p$
interactions by NA22 Collaboration.

 The correlation  function   $C_{2}(Q^2)$   was   also
investigated   in $e^{+}e^{-}$-annihilation at 91 GeV
($Z^0$-boson)  by  DELPHI  Collaboration
(see \cite{15}). In the range $Q^2\leq $0.03 GeV$^2$ the
function
$C_{2}(Q^2)$ of DELPHI  ($e^{+}e^{-}$)
and that of UA1 ($p\bar p$) show quite a similar  shape,
see  Fig.  9.
Since there is a rise for smaller $Q^2$ only for the  same
charged
pairs,  BE  correlation  is  evidently  responsible
for   this behaviour.

     However, the $C_2$ of DELPHI is also rising for smaller
$Q^2$  in
the interval $Q^2\geq 0.03$ GeV$^2$ both for same and opposite
pairs, where the  UA1 data  show  a  comparatively  small  rise.
Therefore   in   $e^{+}e^{-}$
annihilation some other mechanism must be responsible  for  this
power law behaviour which is manifest even in oppositely charged
pairs. Jet evolution or hadronization may play the role.

               \subsection { Correlation integrals.}

     A further insight into the problem of correlations would be
provided by investigation of cumulant moments $K_{q}(\delta )$
describing true
multiparticle correlations of the  order  $q$  with  lower  order
contributions subtracted (cumulants vanish whenever one of  the
particles involved is statistically independent of the  others).
The cumulants relevant to intermittency study are integrals over
(connected) correlation  functions  taken  in  decreasing  phase
space volumes. An experimental investigation of the correlation
integrals is difficult because they suffer from low statistics.
It was performed recently using  star  topology  confinement  in
correlation integrals and $q_{ij}^{2}$ as a "distance" between
particles, that  is  inserting  the  factor  (\ref{2.7})  into
the   correlation
integrals.

     Results  on  correlation  integral  $K_{3}(Q^2)$  for   all
 charged particles and negatives only in $\mu N$  interactions
\cite{13}  are
shown in  Fig.  10a.  The  behaviour  of   $K_{3}(Q^2)$ is
approximately
power-like in $Q^2$ variable. In order to find the  origin
of  the
three-particle correlations in Fig.  10a,  $\mu N$  Monte
Carlo  (MC)
events were generated according to the  Lund  model  without
BE correlations. The MC predictions for $K_{3}(Q^2)$ are shown
in  Fig. 10b.
For (-- -- --) triplets, $K_{3}(Q^2)$ of the MC events is rather
independent of $1/Q^2$ in contrast to the data. This shows that
the rise of
$K_{3}(Q^2)$ in  the data is very likely due to three-particle
BE correlations
which were not incorporated into the Lund MC used.
For $(ccc)$-triplets the
situation  is more  complicated  since  in  the  data  both  BE
correlations (weaker  than in  (-- -- --))  and   resonance
decays contribute. In the MC
(without BE but with resonance decays)  $K_{3}(Q^2)$ is smaller
than in the data but rises due to resonance decays.

     Irreducible higher-order correlations were also established
up to fifth order in  multiparticle  production  in  $\pi ^{+}p$
  and  $K^{+}p$ collisions at 250 GeV/c by EHS/NA22 Collaboration
 \cite{16}.  The star integral method has provided a clear
improvement  over  the
earlier analysis based on the same data. The  charge  dependence
of the correlation was studied in a comparison
of like-charged (Fig. 11a) and unlike-charged (Fig. 11b) particle
combinations. Both charge  combinations  show  non-zero  genuine
higher order correlations and an  increase  of  the  correlation
functions with decreasing interval $Q^2$.

     The correlations  among  unlike-charged  combinations  (i.e.
combinations to  which  resonances  contribute)  are  relatively
strong near $Q^2\sim $1 GeV$^2$ but the increase for  smaller
$Q^2$ is  relatively
slow. Correlations among like-charged particles are  small  at
$Q^2\sim $1 GeV$^2$ but increase rapidly to reach, or
 even cross, those  of  the
unlike-charged combinations at lower $Q^2$. This   difference
diminishes with increasing order $q$. The effective slopes of the
power-like scaling  law  for  the  various  charge  combinations
fitted in the range $0\leq \ln (1$GeV$^2/Q^2)\leq 5$ are given
in the Table 7.

\begin{table}
\begin{center}
\caption{Slopes $\phi _q$ fitted to the
dependence $K_q\sim (Q^2)^{-\phi _q}$}
\begin{tabular}{l|l|l|l|l}   \hline
  charge    &    q = 2       &     q = 3     &
   q = 4      &   q = 5  \\
combination &                &               &
              &         \\ \hline
   all     &$0.205\pm 0.005$ & $ 0.72\pm 0.03$ &
  $1.2\pm 0.2$  &  $2.0\pm 1.0$
\\ \hline
     --~-- &$0.387\pm 0.009$&               &
              & \\
like-charged& & $1.03\pm 0.08$ &  $1.8\pm 0.3$ &   \\
        ++ &$0.438\pm 0.010$& & &    \\ \hline
unlike-     &$0.096\pm 0.004$ & $ 0.61\pm 0.03$ &
  $2.0\pm 0.5$ & \\
charged     &                &               &
              &    \\ \hline
\end{tabular}
\end{center}
\end{table}
One can see that like-charged particles show faster growth of the
     correlation integrals with decreasing $Q^2$-intervals  thus
indicating the
important role of  irreducible  higher order  BE correlations.

            \subsection { Summary of experiment.}

     Correlations  between  different  charge  combinations   in
multiple  production  processes  were  recently  measured  in  a
variety of reactions ($p\bar p, \mu N, e^{+}e^{-}$,\\
$ \pi ^{+}p, K^{+}p$).
The  data  on  two-  and  many-particle
correlations taken as a  function  of  permissible  phase  space
volume  were   presented   (intermittency   study).  This   new
development became possible due to investigation of the  density
and correlation integrals which give a possibility  to  use  the
available  statistics  in  an  optimal  way  and  introduce  any
convenient variable.

     {}From this analysis it becomes clear that  the  increase  of
correlations with decreasing 3-dimensional phase space volume is
essentially due to correlations between like-charged  particles.
An evident candidate for these like-sign correlations is the  BE
statistics.

     The intermittency effect (in its wide sense) depends
strongly on variable chosen. It is poorly seen in
rapidity-analysis and much
more pronounced in three-dimensional-  and $Q^2$-analyses.
The last
variable is one of the most popular nowadays but it  brings  its
own problems as will be discussed below. As we show, the  effect
in higher dimensions becomes stronger just due  to  the  trivial
decrease of the available phase space  (smaller  denominator  in
normalized factorial moments) while the steep rise at smaller
$Q^2$ is determined by the peaked contribution of BE  correlations
  as
exposed in that variable.

    \section { Correlation studies in different variables.}

     \subsection  {     Variables and windows.}

     For  a  proper  interpretation   of   the   intensity   and
correlation integrals $F_{q}^{\Omega }$ and $K_{q}^{\Omega }$ one
has to look  more  carefully
on their structure. In particular it  is  important  to  realize
what kind of windows they provide.

          In general, the density and correlation functions
 $\rho _{q}(p_1,\ldots,p_q)$ and \\ $C_{q}(p_1,\ldots,p_q)$
depend on  $3q$  independent  variables,  say,  $3(q-1)$
independent momentum differences ${\bf q}_{ij}={\bf p}_{i}-
{\bf p}_{j}$
and three components  of  their  average
momentum ${\bf p}=\frac {1}{q}\sum _{j=1}^{q}{\bf p}_{j}$.
The $3(q-1)$ differences must  be  confined  in  some  or
other way to provide an elementary cell in the phase space
(the window) and the remaining 3 variables show the position
  of  the window. The dependence on the cell size is  the
object  of  the intermittency study. The dependence on the
cell  position  is  asubject of an  averaging.  The  last
is  necessary  to  get  an
acceptable statistics. This averaging is explicit in
traditional
(vertical) factorial moments (\ref{1.3}) and implicit in  the
density
and correlation integrals (\ref{1.18}), (\ref{1.19}). Let us
note  that  the
cell size must be independent  of  the  order  of  the  moments;
otherwise they  will  not  be  the  (averaged)  moments  of  any
distribution.

    The choice of appropriate variables is of great  importance
in correlation study. As it could be seen from the  experimental
survey the intermittency  effect  is  poorly  seen  in  rapidity
variable  and  it  is  clearly  seen  in  3  dimensions  and  in
$Q^2$-variable. In BE correlations  one is led to consider the
differences of  three-momenta  of  the  particles
${\bf q}_{ij}={\bf p}_{i}-{\bf p}_{j}$  as  input
variables because the strength of BE effect is  determined  just
by these differences. Considering other  variables  one  has  to
translate the BE correlation  from  ${\bf q}_{ij}$  to
these  variables  to compare the effect with the data.

               \subsection { $Q^2$-confinement.}

     Let us consider in more details the  $Q^2$-variable which is
intensively used in the experimental study of the intermittency.
The  "distance"  between  two  particles  is  defined   now   as
\begin{equation}
q_{ij}^{2}=({\bf p}_{i}-{\bf p}_{j})^{2}-(E_{i}-E_{j})^{2}=
{\bf q}^{2}-q_{0}^{2}
,
\label{3.1}
\end{equation}
and the factor confining permissible phase-space volume contains
the corresponding (step-) $\Theta $-function; $\Theta
(Q^2 -q_{ij}^{2})$. It is not  difficult  to  see
that the above $\Theta $-function confines an ellipsoid
\begin{equation}
{\bf q}_{t}^{2}+\frac {M^2}{E^2}q_{l}^{2}=Q^2 ,    \label{3.2}
\end{equation}
where $M$ is the invariant mass, $M^2 =Q^2 +
4m_{\pi }^{2}$, and
$E$ is the  total  energy
of the pair. So the confined volume in momentum space is
\begin{equation}
V=\frac {4\pi }{3}Q^{3}\frac {E}{M}  .    \label{3.3}
\end{equation}
The longitudinal phase space (along the direction of
the  total
momentum ${\bf p}_{i}+{\bf p}_{j}$) is rising here with
$E$ and $Q$,
\begin{equation}
L_{l}=\frac {QE}{M} .  \label{3.4}
\end{equation}
     On the one hand the $E$-dependence  of  $V$  means  that
the $Q^2$-confined cells (bins) have different sizes in momentum
space
being much larger for  fast  particles.  So  with  $Q^2$ fixed
one
already has an  averaging  over  momentum  cell  sizes.  If  the
correlation  functions  have  the  momentum  difference   as   a
characteristic scale (as it is the  case  for  BE  correlations)
then  the  above  feature  is  rather  unpleasant  because   the
intermittency study suggests an  investigation  of  correlations
taken as a function of the cell size.

     On the other  hand  the  particles  in  increased  momentum
intervals may well become uncorrelated  and  the  integral  over
their correlation function is already saturated  at  some  fixed
momentum scale whereas the cell size $L_l$ is still rising with
 $Q$ due to large $E/M$ factor. In this case the normalized
correlation
integral  will  decrease  with  increasing  $Q^2$  (increase with
decreasing $Q^2$) approximately according to the power law in some
$Q^2$ interval. Just this mechanism was found responsible \cite{17}
for the steep rise of the two-particle correlation  function  of
the like-charged particles with decreasing $Q^2$. As a  conclusion,
$Q^2$ variable is not an  appropriate  variable  for  the  physical
interpretation of the intermittency effect  if  the  correlation
between  particles  has  a  characteristic  scale  in   momentum
difference ${\bf q}_{ij}^{2}$ as it is the case for BE correlations
(at the same
time $Q^2$ being related to  invariant  mass  squared,
$Q^{2}=M^{2}-4m_{\pi }^{2}$,  is  an
appropriate variable for correlations arising due  to  resonance
decay). The steep  and  approximately  power  law  rise  of  the
normalized correlation integrals with  decreasing  $Q^2$ does not
necessarily reflect the corresponding behaviour of the  original
correlation function $C_{q}(p_1,...,p_q)$, being a kinematical
effect inherent in $Q^2$ variable.

       \subsection {Three-momentum difference confinement.}

     In general, different mechanisms responsible  for  particle
correlation bring their own natural variables (the  variable  is
"natural" if the correlation function $C_q$ has a scale parameter
related to this variable). If one considers BE correlations then
the natural variable is three-momentum difference
${\bf q}_{ij}={\bf p}_{i}-{\bf p}_{j}$ (the energy
difference is also involved for nonstationary  particle  sources
and a modification of the variables is necessary  for  expanding
sources). The correlation function  $C_{q}(p_1,\ldots,p_q)$
depends  also  on total
momentum ${\bf P}=\sum _{j=1}^{q}{\bf p}_{j}$ (or on average
momentum
${\bf p}={\bf P}/q$).

     The essential point is  that  these  two  dependences  have
different   momentum   scales:   the   characteristic   momentum
difference scale $p_{d}\sim 1/R$ ($R$  is  an  effective  source
  size  for  BE
correlations) is  noticeably  smaller  than  the  characteristic
transverse and longitudinal momentum  scales  $p_t$  and  $p_l$.
 This
means that the correlation function is nonzero in  a  "tube"  in
$3q$-dimensional space with its axis directed along  the  line
${\bf p}_{1}=\ldots ={\bf p}_{q}$
and with cross-section of the tube of the order $p_{d}^{3(q-1)}$.
 The  length
of the tube is of the order $p_l$ in longitudinal (along the beam
direction) and $p_t$ in transverse direction ($p_l\gg p_t$ in
experiment).

     Let us now confine the 3-momentum differences imposing  the
condition  $\vert {\bf q}_{ij}\vert \leq \delta $  (the  window).
The  result  for  the  normalized
correlation integral $K_q$ (see (\ref{1.19})) depends strongly
on the
relationship of $\delta $ and the scale parameters
$p_d\ll p_t\ll p_l $.

     a) The region of  very  small  $\delta <p_d$,  where  the
correlation
function is maximal, is not still accessible in experiments.

     b)  In  the  region  $p_d < \delta < p_t$  the  integration
  over momentum
differences  in  $K_q$  is  saturated  with  the  scale   $p_d$.
  All corresponding integrations are cut by $\delta $  in  the
normalization
factor $N_q$. As a result, we are left with  a  dependence  of the
form
\begin{equation}
K_q = \frac {K_{q}^{(u)}}{N_q}\sim \left (\frac {p_{d}^{3}}
{\delta ^3}\right )
^{q-1} .   \label{3.5}
\end{equation}

          c) In the region $p_t <\delta <p_l$  only
integrations  over
longitudinal
momenta are cut off and we have a dependence
\begin{equation}
K_q \sim \left (\frac {p_{d}^{3}}{p_{t}^{2}\delta }
\right )^{q-1} . \label{3.6}
\end{equation}

          d) In the full phase-space ($\delta \sim p_l$) the
normalized correlation
integral takes its minimal value
\begin{equation}
K_q \sim \left (\frac {p_{d}^{3}}{p_{t}^{2}p_{l}}\right )^{q-1} .
 \label{3.7}
\end{equation}
     The  above  rough  estimation  shows  that  the  normalized
correlation integral obeys approximately power-law $\delta
$-dependence
in some subintervals of the window width $\delta $. The
dependence  is
steeper for small windows (the effective intermittency index in
the region b) is three times larger than in the region c)) and
for higher order of the moments.

     We conclude that the intermittency  effect  (in  the  broad
sense  of  the  word)  is  strong  and   ensures   approximately
power-like behaviour in three-dimensional analysis  in  momentum
difference  variables.  No  specific  physical  phenomena   were
involved to get  the  above  power-law  behaviour.  It  is  just
connected  with  the  trivial  dependence  of  moments  on   the
available  phase  space  and  does  not  involve  any  dynamical
"anomalous" dimension as it is the case for QCD  jets  discussed
in the next section. The transition from "$q^3$-regime" in
(\ref{3.5}) to
"$q$-regime" in (\ref{3.6}) is also  due  to  the  transverse
momentum
limitation  which  is  an  inherent  property  of  multiparticle
production. Actually, the dependence on $\delta $ is mostly
determined by  the
"kinematical" normalization  factor  in  the  denominator  which
shows how large is the average  multiplicity  within  the  phase
space window provided by the confinement condition.

     \subsection { An illustrative example (BE correlations).}

     To get somewhat more detailed picture of the intermittency
effect
induced by BE correlations let us consider the cumulant  moments
for BE correlations in a simple environment. We suggest that the
particles  are  created  by  some   random   gaussian   currents
\cite{18a}.
( A generating functional for BE correlations with chaotic and
coherent components is formulated in Ref.\cite{18b}.)
Then the particle densities $\rho _{q}(p_1,\ldots,p_q)$  and  the
correlation functions $C_{q}(p_1,\ldots,p_q)$ can be
expressed through a single quantity $F(p_{i},p_{j})$ which
is an averaged current correlator. In particular
\begin{equation}
\rho _{1}(p)=F(p,p) , \label{3.8}
\end{equation}
and the (unnormalized) correlation integral (\ref{1.19}) is
\begin{equation}
K_{q}^{(u)}(\delta )=(q-1)!\prod _{j=1}^{q}\int dp_{j}
F(p_1,p_2)F(p_2,p_3)
\ldots F(p_q,p_1)\Omega _{q}(\delta ;p_1,\ldots,p_q) .
\label{3.9}
\end{equation}
We represent $F$-function in the form
\begin{equation}
F(p_1,p_2)=[\rho _{1}(p_1)\rho _{1}(p_2)]^{1/2}d_{12}(p_1-p_2) ,
\;\; d(0)=1
 \label{3.10}
\end{equation}
suggesting that the function $d_{ij}$ depends  only  on  the
momentum
difference. Remembering that the correlation scale  $p_d$
inherent
in $d_{ij}$ is much smaller than characteristic particle
momenta  $p_t, p_l$,
we take the densities $\rho _{1}(p_j)$ in Eq. (\ref{3.9})
in  coinciding points  to get
\begin{equation}
K_{q}^{(u)}(\delta )=(q-1)!\int dp\rho _{1}^{q}(p)
\prod _{j=1}^{q-1}d(p_j
-p_{j+1})
d_{12}\ldots d_{q1}\Omega _{q}(\delta ;p_1,\ldots ,p_q)
  \label{3.11}
\end{equation}
with $q-1$-fold integration over differences $p_j -p_{j+1}$.

     Being interested in qualitative results we may  cancel  one
of  $d$-functions  in  Eq.  (\ref{3.11})  taking
$d_{q1}=d_{q1}(0)=1$.
This  leads   to
inessential  numerical   misrepresentation   of   the   integral
retaining  its  qualitative  behaviour   because   one   of   $q$
$d$-functions in (\ref{3.11}) does not serve as a direct
cut  factor.
On an equal footing we may neglect a variation  of  one-particle
density $\rho _{1}(p)$ in the region where the density is
substantial,  this
region being confined by  $p_t$  and  $p_l$  as  above.
The  momentum
difference window may be defined by a confinement of  successive
momentum differences,
\begin{equation}
\Omega _{q}(\delta )=\Theta (\delta -q_{12})\Theta
(\delta -q_{23})\ldots
\Theta (\delta -q_{q-1,q})     \label{3.12}
\end{equation}
(the snake topology).
     As a result, the correlation integral (\ref{3.11}) takes
a simple form
\begin{equation}
K_{q}^{(u)}(\delta )\approx (q-1)!\int ^{(p_t, p_l)}dp
\rho _{1}^{q}
(\int _{\delta }dq d(q))^{q-1} .    \label{3.13}
\end{equation}
The normalization factor $N_q$  in  Eq. (\ref{1.19})  taken  in
the  same
approximation is equivalent to $\langle n\rangle ^q$,
\begin{equation}
N_q \approx \int ^{(p_t, p_l)}dp\rho _{1}^{q}
(\int _{\delta }^{(p_t, p_l)}
dq)^{q-1} \approx \langle n\rangle ^q ,   \label{3.14}
\end{equation}
and the correlation integral (\ref{3.13}) can be written
in the form
\begin{equation}
K_{q}^{(u)}(\delta ) \approx (q-1)!\langle n\rangle ^{q} k^{1-q}
\label{3.15}
\end{equation}
with
\begin{equation}
k=\frac {\int _{\delta }^{(p_t, p_l)}dp}{\int _{\delta }dp d(p)}
\approx
\frac {\int ^{\min (p_t,\delta )}d^{2}p_{t}
\int ^{\min (p_l,\delta )}dp_{l}}
{\int ^{\min (p_d,\delta )}d^{3}p}\geq 1 .   \label{3.16}
\end{equation}
     The  cumulant  moments  (\ref{3.15})  correspond  to
generating
function (\ref{1.12}) of the form
\begin{equation}
\ln g(z)=-k\ln (1-\frac {\langle n\rangle z}{k}) .   \label{3.17}
\end{equation}
This  is  the  generating  function  of  the  negative  binomial
distribution (NBD)
\begin{equation}
P_n = \frac {\Gamma (n+k)\Gamma (k)}{\Gamma (n+1)}
(1+\frac {\langle n\rangle }
{k})^{-k}(1+\frac {k}{\langle n\rangle })^{-n} .    \label{3.18}
\end{equation}
(let us note that the linked pair approximation  also  leads  to
NBD \cite{19}). The parameter  "$k$"  of  NBD  is  given  by  Eq.
(\ref{3.16}). It decreases with decreasing of permissible
interval $\delta $.
In a rough step function approximation it reduces to
\begin{eqnarray}
k=1 \;\;\;\; for \;\; \delta <p_d ,\\
k\approx \delta ^{3}/p_{d}^{3}  \;\;  for \;\; p_d <
\delta <p_t ,\\
k\approx p_{t}^{2}\delta /p_{d}^{3}  \;\; for \;\; p_{t} <
\delta <p_l ,\\
k\approx p_{t}^{2}p_{l}/p_{d}^{3}  \;\; for  \;\; \delta \sim p_l
\label{3.19}
\end{eqnarray}
in an accordance with estimations (\ref{3.5})-(\ref{3.7}) of the
normalized correlation integral.

    {}From the above example one can  clearly  see  the  role  of
BE-correlations in the intermittency  effect.  For  very  narrow
windows  ($\delta \rightarrow 0, k \rightarrow 1$)  we   have   a
geometrical   distribution   (GD)
characteristic for thermal ("totally chaotic") excitation of the
particle source. This distribution, having  normalized  cumulant
moments $K_q =(q-1)!$ and normalized factorial moments $F_q =q!$,
is  the widest
(having maximal moments) distribution possible  in  the  present
scheme, where BE correlations were estimated for a single  fixed
scale particle source. To get  larger  values  of  moments  (for
experimental evidence see Fig. 3, Ref. \cite{10})  one  has  to
introduce additional fluctuations  or  to  revise  the  particle
source form.

     For wide enough windows ($\delta \rightarrow p_l$) one  has
 $k\gg 1$  leading  to  the
relatively narrow Poisson distribution ($PD=\lim _{k\rightarrow
\infty }NBD$). This means that  BE
correlations which we consider here are not effective  in  large
phase space volume. The change of the window width from zero  to
infinity interpolates between GD and PD and this  is  a  general
feature of the present scheme. NBD is  one,  especially  simple,
kind of the interpolation and it  is  not  surprising  that  NBD
appears as a possible approximation. More  accurate  estimations
of the correlation integral (\ref{3.9}) lead to  similar
conclusions. Additional coefficients $\gamma _{q}$ appear in the
cumulant moments
(\ref{3.15})
and a single parameter $k$ varies slightly for  different  orders
$q$,  but  the  qualitative  behaviour  (\ref{3.15}),  (\ref{3.16})
   remains unchanged.

    \subsection         {The (pseudo) rapidity confinement.}

     Let us turn now to the rapidity analysis  which  initiated
 the
whole story of intermittency in particle physics. We  show  that
pseudorapidity (which coincides with rapidity  for  relativistic
particles) is the most suitable variable to  study  correlations
providing jet-like structure of high  energy  events.  Moreover,
such events give rise to (quasi)intermittent (in strict  sense!)
behaviour of correlations in this variable.

     Really, we will speak  about  the  one-dimensional  angular
confinement when the bin of a definite size $\delta \theta $ in a
polar  angle
is chosen. The pseudorapidity size $\delta \eta $ is proportional
to $\delta \theta$
\begin{equation}
\delta \equiv \delta \eta \approx \delta \theta /\theta
\label{3.20}
\end{equation}
with a weight given by a location $\theta $ of the center  of
the  bin
that is trivially accounted when averaging over all locations is
done. The dimension of the analyzed bin enters the results in  a
very simple (even trivial) manner as it is shown below.

     Let us consider in QCD the correlation within the gluon jet
emitted    by    a    quark    produced    in
$e^{+}e^{-}$-annihilation
\cite{20,21,22,23}. We are interested in correlations among  the
partons (mostly gluons) created  during  the  evolution  of  the
initial gluon and fitting the pseudorapidity bin $\delta \eta $.
They belong to some subjet inside the primary jet which is
separated  from
others  due  to  the  angular  ordering  in  QCD.  Surely,   the
prehistory of a jet as a whole  is  important  for  the  subjet
under consideration as is shown in Fig. 12.

     Here
\begin{enumerate}
\item the primary quark emits the hard gluon with energy  $E$  in
the direction of the angular interval $\delta $,  but  not
necessarily hitting the window,
\item the emitted gluon produces the jet  of partons with parton
splitting  angles  larger  than  the  window size,
\item among those partons there exists such a parton (subjet)
with energy $k$ which hits the window,
\item all decay products of that parton  subjet  cover  exactly
the bin $\delta $.
\end{enumerate}
This picture  dictates  the  rules  of  calculation  of  the
$q$-th correlator  of  the  whole  jet.  One  should  average
the   $q$-th
correlator of the subjet $F_{q}^{(u)}$  over  all  possible
ways  of  its
production i.e. convolute it with the inclusive spectra of
such partons $D^{(\delta )}$ in the whole jet and with the
probability of creation
of the jet $\alpha _{S}K_{F}^{G}$. Analytically, it is
represented by \begin{equation}
F_{q}^{(u)}(E_{0}\delta )\sim \int ^{E_{0}}\frac {dE}{E}
\frac {\alpha _S}{2\pi }
K_{F}^{G}(E/E_{0})  \int ^{E}\frac {dk}{k}D^{(\delta )}
F_{q}^{(u)}(k\delta ) ,
\label{3.21}
\end{equation}
where $E_{0}$ is the primary energy, $E$ is the jet energy,
and $k$ is the energy of the subjet hitting the window.
Since the unnormalized moments increase with  energy  while  the
parton spectrum decreases, the product $D^{(\delta )}(k)
F_{q}^{(u)}(k\delta )$
has a maximum at  some
energy $k_{\max }$, and the integral over momenta may  be
calculated  by
the steepest  descent  method.  Leaving  aside  the
details  of
calculations (see \cite{22}), we describe the general
structure
of the correlator for the fixed coupling constant
$\gamma _{0}=(6\alpha _{S}
/\pi )^{1/2}=\mbox{const}$
\begin{equation}
F_{q}^{(u)}(\delta )\sim \Delta \Omega
(\delta )^{-\gamma _{0}/q}
(\delta )^{q\gamma _{0}} ,  \label{3.22}
\end{equation}
where the three factors represent the phase  space  volume,
the energy  spectrum  factor  and  the  $q$-th  power  of
the   average
multiplicity. To get the  normalized  factorial  moment
$F_{q}(\delta )$  one
should divide (\ref{3.22}) by the $q$-th power of that part
of the mean
multiplicity of the whole jet which appears inside the
window $\delta $
i.e.  by  the  share   of   the   total   average   multiplicity
corresponding to the phase space volume $\Delta \Omega $:
\begin{equation}
\Delta n(\delta )\sim \Delta \Omega \langle n\rangle .
\label{3.23}
\end{equation}
If the analysis has been done in the $d$-dimensional  space,
 the phase space volume is proportional to
\begin{equation}
\Delta \Omega \sim \delta ^d ,  \label{3.24}
\end{equation}
where  $\delta $  corresponds  to  the  minimal  linear  size
of   the $d$-dimensional window.The last statement stems from
the singular behaviour  of
parton propagators in quantum chromodynamics (see \cite{22}).
That is why the factorial moments may be represented as  products
 of the purely kinematical factor depending on the dimension
of  the
analysed space and of the dynamical factor which is not  related
to the dimension and defined by the coupling constant
\begin{equation}
F_{q}\sim (\delta )^{-d(q-1)}(\delta )^{(q^{2}-1)\gamma _{0}/q} .
\label{3.25}
\end{equation}
At small angular windows $\delta $ the intermittency indices
are  given by
\begin{equation}
\phi _{q}=d(q-1)-\frac {q^2 -1}{q}\gamma _{0} . \label{3.26}
\end{equation}
This formula  is only valid  for  moderately  small  bins
when  the
condition \\ $\alpha _{S}\ln (\Delta /\delta )<1$ is fulfilled.
For  extremely  small  windows,  one
should take into account  that  the  QCD  coupling  constant  is
running.  Then  the  constant  $\gamma _{0}$  should  be
replaced by  the
effective value $\langle \gamma \rangle $ which depends
logarithmically on the width of
the window $\delta $. As a result (see \cite{22}), numerical
values  of
the intermittency indices for very small bins become  noticeably
smaller than in the fixed coupling regime,  especially  for  the
low-rank moments. Moreover, the simple  power-law  behaviour  is
modified  by  the   logarithmic   correction   terms   and   the
intermittency indices  depend  on  the  value  of  the  interval
chosen. The resulting curve of $\ln F_{q}(\delta )$ as a function
of $-\ln \delta $  has  two
branches and qualitatively reminds those shown in  Fig.  1.  The
rather steep linear increase at the moderately small  bins  with
the slope (\ref{3.26}) is  replaced  at  smaller  windows
by  much slower quasi-linear  increase.  It  is  easy  to
calculate  the
location of the transition point to another regime and  to  show
that at higher values of $q$ it shifts to  smaller  bin  sizes
in accordance with trends in Fig. 1. We have described the
results
of the double logarithmic  approximation  of  QCD.  Higher-order
terms have been treated in \cite{22}.  They  do  not  spoil  the
general conclusions providing the corrections of the order of 10
per cents.

     It is interesting to note that  higher  dimension  analysis
just  adds  an  integer  number  to  the  trivial  part  of  the
intermittency  index  and  does  not  change   its   non-trivial
"anomalous"  dimension.  Thus  the  increase  of   intermittency
indices in higher dimension is trivial and  has  nothing  to  do
with jet dynamics but is a consequence of  phase  space  factors
(mostly  in  the  normalization  denominator).   The   important
difference from BE-effect is that the numerator of  the  moments
provides the non-trivial "anomalous" part of  the  intermittency
index which is absent in  BE-treatment.  It  is  common  in  all
dimensions.

     The above results may be restated in terms of fractals. The
power-like behaviour of factorial moments points out to  fractal
properties of particles (partons)  distributions  in  the  phase
space. According to the general theory of fractals (see \cite{1}
and references therein), the intermittency indices  are  related
to fractal (Renyi) dimensions $D(q)$ by the formula
\begin{equation}
\phi _{q}=(q-1)(d-D(q)) ,  \label{3.27}
\end{equation}
wherefrom one gets (see (\ref{3.26})):
\begin{equation}
D(q)=\frac {q+1}{q}\gamma _{0}=\gamma _{0}+\frac {\gamma _{0}}{q} .
\label{3.28}
\end{equation}
The first term corresponds to monofractal behaviour and  is  due
to the average multiplicity increase. The  second  one  provides
multifractal properties and is related to  the  descent  of  the
energy spectrum as discussed above. It is clearly seen that  the
fractality in  quantum chromodynamics  has  a  purely  dynamical
origin $( D(q)\sim \gamma _{0} )$ related to the cascade nature of
the process while the
kinematical factor in  (\ref{3.27})  has  an  integer  dimension.
 The attempts to relate the fractal properties in the momentum
space to the fractal structure of colliding objects  in  the
ordinary
space were tried also (\cite{24}). The  fractality  in  momentum
space can be also  formulated  as  the  fractal  nature  of  the
subsequent available phase space at each branching of the  gluon
jet (\cite{25,26}).

     Coming back to our problem we would like to stress that the
angular  variable  is  the  most  convenient  one   to   analyse
correlations originated by  the  jet-like  structure.  At  first
sight the opening angle of the jet seems even a more  convenient
(and "natural") variable.  However,  it  is  more  difficult  to
incorporate this angle into above theoretical study  than  just
the  usual polar angle $\theta $. It is easy to show that the
relative  angle  of two partons $\theta _{12}$ is connected with
their polar emission angles
$\theta _{1}, \theta _{2}$,
 and with their relative distances in  pseudorapidity
$\eta _{12}$  and
azimuthal angle $\varphi _{12}$ by the formula
\begin{equation}
\theta _{12}^{2}\approx \theta _{1}\theta _{2}(\eta _{12}^{2}+
\varphi _{12}^{2})    \label{3.29}
\end{equation}
at $\theta _{1}\ll 1$, $\theta _{2}\ll 1$. Therefore the analysis
in the relative angle variable
corresponds to two-dimensional analysis  in  $\eta _{12}$  and
$\varphi _{12}$  when  the  size  of the two-dimensional
interval depends on  polar  angles  of  emitted  partons  and
should be larger at small polar angles. It would produce  higher
average multiplicities in the denominator of  factorial  moments
and suppress their values  at  small  relative  angles  what  is
observed in experiment \cite{12}.

          It exemplifies our statement about "natural" variables
 for each mechanism responsible  for  correlations  in
multiparticle production.

                 \section {Discussion and conclusions.}

     The very  first  intermittency  studies  were  aimed  at  a
scaling law in behaviour of factorial moments with hope to  find
out new collective effects in high energy interactions. Later it
was recognized that it is just a part of the day-to-day work  on
correlation properties of multiparticle production that does not
diminish the importance of the above problem but helps  also  to
disentangle  contributions  of  different  known  mechanisms  to
particle correlations. According to our present-day  theoretical
prejudices we can name at least four of  them.  At  the  initial
stage  the  quark-gluon  jets  appear.  If  described   by   the
perturbative  QCD  they  should  give  rise   to   the   (quasi)
intermittent behaviour of factorial moments as functions of  the
(pseudo)rapidity  bin  size.  At  a  simplified  level  of   the
hypothesis on local parton-hadron duality it should be valid for
final hadrons as well. However, the transition from  partons  to
hadrons could be not as simple as  that,  and  it  is  sometimes
considered as a  phase  transition  imposing  its  own  critical
indices. In addition to it, the final stage interactions  giving
rise to known resonant states, surely, play important role.  The
final stage of creating hadrons  asks  for  symmetry  properties
such as Bose-Einstein symmetrization to be respected. If one  is
interested  in  looking  for  new  non-traditional  sources   of
correlations (stochasticity,  instabilities  etc.)  one  should,
first, to show that they contribute to correlations  differently
compared to considered "traditional" sources.

     As we mentioned above, each of them should be described  in
its own natural variable connected to its characteristic  scale.
We tried to show that resonances bring with them the mass  scale
of squared 4-momenta, BE-symmetrization is  better  revealed  in
3-dimensional momentum analysis, while  the  jet-like  structure
asks for angular (or (pseudo)rapidity)  variable  with  a  scale
determined by the corresponding "length" of the  shower  related
to the (running) coupling constant.

     When analysed in "unnatural"  variables,  these  mechanisms
can produce the dependences which are not typical for  them  and
mask (or imitate) some other  effects  due  to  the  determinant
 of  the transformation. It   happens,   for   example,   with
BE-contribution when looked in $Q^2$-variable. It becomes
strongly peaked at small $Q^2$ and imitates power-like law. One
should  not
attempt to fit it  by  "traditional"  Gaussian  dependence  when
looked in that variable as well as one should not claim that  it
produces intermittency in a strict sense.

     Unfortunately, that example  demonstrates  also,  that  the
contribution of some mechanism in its  "unnatural"  variable  is
not necessarily smeared out but, on the  contrary,  can  produce
rather steep dependence provoking misleading conclusions.

     The choice of (pseudo)rapidity as a "natural"  variable  in
the original paper \cite{2} was just  related  to  traditions  in
theoretical approach and to  experimental  facilities.  It
appeared  to  be natural for jets but  not  for
resonances  and  BE-correlations
which seem to be  smeared out in that variable.  Separate  study
of  the  quantitative  contributions  of  different   mechanisms
plotted as functions of the same variable asks  for  Monte-Carlo
calculations. However, main  qualitative  ingredients  are  seen
from above analytical approaches.

     In  our  opinion,  the  strong  "kinematical"  phase  space
dependence provided by the denominator of the normalized moments
spoils the analysis introducing strongly  increasing  (at  small
$\delta $) factor depending on the dimension of the analysed bins.
  To unify the intermittency indices it looks  appropriate
to  leave
just  $\langle n\rangle $  in  the  denominator  of  moments  that
cancels   all
kinematical  factors  at  the  expense  of  introducing   energy
dependence.

          To conclude, we  have  shown  that  different  effects
are better displayed if their "natural" variables  are  chosen.
Some proposals are discussed above, but that study is just at the
very initial stage and we call for further  elaborated  criteria
beside those considered in the present paper. What concerns the
title of the paper, we can say that Bose-Einstein correlations
do  contribute
to intermittency in a wide sense. However, intermittency in  its
initial meaning exists even  for  unlike-charged  particles  and
should be ascribed (probably) to jet-like parton cascades but not
to Bose-Einstein correlations.

\vspace{0.5cm}
\begin{center}
{\large Acknowledgements}
\end{center}
This work is partially supported by the JSPS
Program on Japan-FSU Scientists Collaboration.
Andreev and Dremin are supported by Russian fund for
fundamental research (grant 93-02-3815). Moreover, Biyajima
 is partially supported by Japanese Grant-in-Aid for Scientific
Research from the Ministry of Education, Science and Culture
(No. 06640383). Suzuki is grateful for the financial support
by Matsusho Gakuen Junior College in 1994.
\newpage
\vspace{0.5cm}
\begin{center}
                      { \large         Figure captions}
\end{center}
\vspace{0.5cm}
\begin{tabular}{l|l}
Fig. 1. &Factorial moments  of  order  $q=2, 3, 4$  for  the
all  charged\\
        &sample, and the restriction  to  the
positive-only  and\\
        &negative-only samples in $\pi ^{+}p$ and $K^{+}p$
interactions \cite{8}.\\
Fig. 2. &The rise of the factorial moments and density
integrals\\
        &with decreasing $\delta \eta $ in $p\bar p$
collisions \cite{9}.      \\
Fig. 3. &Log-log plot of the second factorial moment
$F_2$  in  one \\
        &dimension ($a, b$) and three dimensions ($c, d$)
for  unlike  ($a,
c$)\\
        &and negative ($b, d$) charges. The full dots show
the  data,      \\
        &the open circles are Lund predictions \cite{10}. \\
Fig. 4. &Comparison of a), b) conventional factorial
moments  $F_2$       \\
        &and $F_3$ in three dimensions to c),
d) $F_{2}^{GHP}$ and         \\
        &$F_{3}^{GHP}$  obtained
        from density  integral  method  in
GHP-topology.\\  &Solid
        lines in a) and c) correspond to fits
according to (\ref{2.1})  \cite{8}.      \\
Fig. 5. &The $\ln F_{q}^{GHP}$ as function of
$-\ln (Q^2/1$GeV$^2)$. \\
        &Note that the abscissa value 0.65  corresponds  to
        the peak of\\ & $\rho $-meson and 1.77 is the
value corresponding
        to the $K_{S}^{0}$ mass \cite{8}.
\\
Fig. 6. &Log-log plot of a) $F_{2}(Q^2)$ for (~--~--~)
and (~+~--~)  pairs
and\\ &$F_{3}(Q^2)$  for (~--~--~--~) and ($ccc$)
triplets vs $1/Q^2$ \cite{13}.  \\
        Fig. 7.  &The rise of the density integrals with
decreasing $Q^2$  in
$p\bar
        p$ collisions \cite{9}.                        \\
        Fig.
        8.  &The  normalized  two-body
correlation  function   for       \\
&different charge combinations a) as a function of
$\delta \eta $, b) \\ &as a
        function of $Q^2$ \cite{9}.     \\
Fig. 9. &A comparison of $p\bar p$ collider data
(UA1) with $e^{+}e^{-}$ at the
        $Z^0$                    \\
&pole (DELPHI) for the second order correlation function \cite{15}.\\
Fig.10. &Log-log
plot of $K_{3}(Q^2)$ for (~--~--~--~) and ($ccc$) triplets
a)  from   \\ &data
        and b) from the Lund Monte Carlo program  including   \\
        &resonances but without BE correlations \cite{13}.    \\
        Fig.11.  &$\ln K_{q}(Q^2)$ as a function of $-\ln Q^2$
\\ &a) for like-charged  and
b)  for unlike-charged particle combinations,\\ &
        compared  to  the expectations from FRITIOF model \cite{16}.
  \\
Fig.12. &The subjet hitting the  window  $\delta \equiv \theta $
originates  from \\ &a parton which appeared in
the  evolution of  the
        primary  \\ &gluon emitted by a parent quark.
        \\ \end{tabular}

\end{document}